\documentclass[]{aa}
\usepackage{lscape,graphicx}
\usepackage{natbib}
\bibpunct{(}{)}{;}{a}{}{ }
\begin{document}

\title{ NGC300 X-1 and  IC10 X-1: a new breed of black hole binary?}
\author{R. Barnard, J. S. Clark, U. C. Kolb}
\offprints{R. Barnard, \email{r.barnard@open.ac.uk}}

\institute{ The Department of Physics and Astronomy, The Open University, Walton Hall, Milton Keynes, MK7 6BT, U.K.}
\date{ Received / Accepted}

\abstract {[ABSTRACT ABRIDGED] IC10 X-1 has recently been confirmed as a black hole (BH) + Wolf-Rayet (WR) X-ray binary, and NGC300 X-1 is thought to be. IC10 X-1 and NGC300 X-1 have  similar X-ray properties, with  luminosities $\sim$10$^{38}$erg s$^{-1}$, and  orbital periods $\sim$30 hr.}
{We investigate similarities between these systems, as well as differences between them and the known Galactic BH binaries.}
{We have examined XMM-Newton observations of NGC300 X-1 and IC10 X-1. For each observation, we extracted lightcurves and spectra; power density spectra  (PDS) were constructed from the lightcurves, and the X-ray emission spectra were modeled.}
{ Each source exhibits PDS  that are characteristic of  turbulence in wind accretion or high state  X-ray binaries (XBs). In this state, Galactic XBs with known BH primaries have soft, thermal emission; however the emission spectra of NGC300 X-1 and IC10 X-1  in the XMM-Newton observations are  predominantly non-thermal.  The remarkable similarity between the behaviour of NGC300 X-1 in Observation 1 and that of IC10 X-1 lends strong evidence for NGC300 X-1 being a BH+WR  binary. }
{The unusual spectra of NGC300 X-1 and IC10 X-1 may be  due to these systems existing in a persistently high state, whereas all known BH LMXBs are transient. BH XBs  in a persistent high state could retain their corona, and hence exhibit a large non-thermal component. LMC X-1 is a  BH XB that has only been observed in the high state, and its spectrum is remarkably similar to those of our targets.  We therefore classify  NGC300 X-1, IC10 X-1 and perhaps LMC X-1 as a new breed of BH XB, defined by their persistently high accretion rates and consequent stable disc configuration and corona. { This scenario may also explain the lack of ultraluminous X-ray sources in the canonical soft state.}}
 
\authorrunning{Barnard et al.}
\titlerunning{ A new breed of black hole binary?}
\maketitle

\section{Introduction}

Characterised by high temperatures and large mass loss rates, Wolf Rayet (WR) stars represent the final stage of stellar
evolution prior to supernova for massive stars \citep[$>$25 M$_{\odot}$;][]{review}. Their spectra are characterised by strong, broad emission lines, which reveal their chemically evolved nature due to, successively, the CNO (WN stars) and 
triple-$\alpha$ (WC stars) processes. These products of nuclear burning become visible as the  hydrogen rich mantle of the
 O star progenitor is stripped away. For single stars this is accomplished via the stellar wind of the progenitor; however 
binary interaction may also contribute to this process. Indeed, if a stellar merger may be avoided, 
interaction between the two components of a high mass X-ray binary (OB star and compact object) is thought to lead to a
 short period WR+compact object binary (WR+co). 

Empirically, such systems  are exceedingly rare; to date we know of only three possible WR+co systems: Cygnus X-3 in our
 Galaxy \citep{lom05}, IC 10 X-1 \citep{cc04, bb04}, and most recently, NGC 300 X-1 \citep{carp07a,cch07}.  Population 
synthesis carried out by \citet{lom05} suggests that $\sim$1 WR + BH binary, and $\sim$1 WR + neutron star (NS) binary,  may exist  as bright X-ray sources in a galaxy like our own; 
this is certainly in keeping with the rarity of candidates.

 \citet{prest07}  re-analysed archival Gemini Multi-Object Spectrograph spectra of the WR star  associated with IC10 X-1, and  found evidence for a  black hole (BH) primary, with  a  plausible BH mass range of 23--34 M$_{\odot}$. This result was subsequently confirmed   by \citet{sf08}, who obtained spectra over five nights of observations with the Keck-I 10m telescope. They report a BH mass of 20--35 M$_{\odot}$ for the compact object,  making it the most massive known stellar BH.  NGC300 X-1 has not yet been optically identified as a WR+BH binary; however, in this work we find striking similarities between the X-ray properties of NGC300 X-1 and the confirmed WR+BH binary IC10 X-1. We will show that NGC300 X-1 is likely to be a WR+BH binary also.

WR+co binaries are just one  subset of high mass X-ray binaries (HMXBs), and it is possible that their temporal and spectral characteristics closely resemble other types of HMXBs. Hence it is instructive to  briefly review the   pertinent properties of HMXBs in general.
HMXBs are well classified by the nature of the donor star \citep[see e.g.][ and references within]{wnp95}.

  Around 60\% of Galactic HMXBs include a Be donor \citep{lvv07}; these have elliptical orbits lasting tens to hundreds of days, only accreting near periastron. As a result, Be HMXBs are transient sources, generally  powered by Bondi-Hoyle accretion \citep{bh44} from a decretion disc around the Be star \citep{on01}.
 Several  Bondi-Hoyle accreting   HMXBs with a  NS accretor  exhibit rapid aperiodic variability in X-ray intensity with amplitude $\sim$10--30\%;  their power density spectra (PDS) resemble a power law with  spectral index $\gamma$ $\sim$1 \citep[{ power $\propto$ $\nu^{-\gamma}$ where $\nu$ is the frequency}, see e.g.][]{nag89,bh90,vdk95}.
  These works suggested that such variability could be present in all accretion-powered pulsars. However,  an inspection of the catalogues of HMXBs in the Galaxy and Magellanic clouds \citep{lvv05,lvv07} revealed that only 16 out of 242 HMXBs have published records of  aperiodic variability.  HMXBs that have exhibited  the variability described above  have power law emission spectra with spectral index, $\Gamma$, $\sim$0.5--1.8.  There are no known BH HMXBs that accrete in this way.

 A further $\sim$32\% of Galactic HMXBs contain supergiant (SG) donors \citep{lvv07}; the orbital periods range over 1.4--41 days, and the donor stars are (or are close to) filling their Roche lobe \citep[see][ for a review]{kap04}. The compact object in SG HMXBs is continuously accreting from the donor's wind and  can be either wind-fed  via Bondi-Hoyle accretion (with luminosities $\sim$10$^{35}$--10$^{36}$ erg s$^{-1}$) or disc-fed, reaching $\sim$10$^{38}$ erg s$^{-1}$  \citep{kap04}. 

Studies of Galactic disc-accreting XBs have shown that their spectral and variability properties depend as much on the accretion rate as the nature of the accretor (NS or BH). At low accretion rates, the emission spectra and temporal variability of disc-accreting binaries are remarkably similar \citep{vdk94}. However, at certain, higher, accretion rates there are characteristic differences that may allow one to determine the nature of the primary; the PDS are well described by a power law with  index $\gamma$ $\sim$1, but the emission spectra for NS and BH XBs are very different. For a disc-accreting XB with a NS primary, the emission corresponding to  the power law PDS is non-thermal, while disc-accreting BH XBs exhibiting the same PDS have a thermal X-ray emission spectrum \citep{vdk94}. Indeed, \citet{don04} examined $\sim$ 1 Tbyte of RXTE archival data on Galactic XBs, and found a region in colour-colour space that is uniquely associated with BH XBs in this high, soft state.  Hence, we may learn a great deal about disc-accreting binaries from their X-ray variability and emission spectra.

\begin{table}[!b]
 \centering
  \caption{ Journal of XMM-Newton observations of NGC300 X-1 and IC10 X-1. For each observation, we provide the target, date, revolution number, exposure and good time interval. }\label{journal}
  \begin{tabular}{cccccc}
  \noalign{\smallskip}
  \hline
  \noalign{\smallskip}
  Object &  Date  & Rev  &  Exp&GT\\
\noalign{\smallskip}
 \hline
\noalign{\smallskip}  
NGC300 X-1  & 2000 Dec 26 & 0192 &  32 ks& 25 ks \\
NGC300 X-1  & 2001 Jan 01 & 0195 & 40 ks & 40 ks \\
NGC300 X-1  & 2005 May 22 & 0998 & 35 ks & 25 ks \\
NGC300 X-1 & 2005 Nov 25 & 1092 & 35 ks & 35 ks \\
IC10 X-1 & 2003 Jul 03 & 0653 & 44 ks & 28 ks \\
\noalign{\smallskip}
\hline
\noalign{\smallskip}
\end{tabular}
\end{table}

 NGC300 X-1 and IC10 X-1 both have orbital periods $\sim$30 hr and X-ray luminosities $\sim$10$^{38}$ erg s$^{-1}$ \citep{carp07a,prest07,sf08}. We examined all existing XMM-Newton observations of NGC300 X-1 and IC10 X-1, in order to compare their properties. In particular, we examined the PDS from the X-ray lightcurves of these systems for the first time. In Sect.~\ref{obs} we describe the observations and data analysis, then provide the results from NGC 300 X-1 and IC10 X-1 in turn in Sect.~\ref{res}. We discuss our findings in Sect.~\ref{discuss},  and consider four scenarios for NGC300 X-1 and IC10 X-1: wind accretion onto a neutron star, disc accretion onto a neutron star, wind accretion onto a black hole and disc accretion onto a black hole. { We then compare the properties of  these sources with those of the ultraluminous X-ray sources (ULXs), many of which are thought to be HMXBs.} Finally we  draw our conclusions in Sect.~\ref{conc}.

\section{Observations and data analysis}
\label{obs}

Four XMM-Newton observations have been made of NGC 300 X-1, and one of IC10 X-1; a journal of observations is provided in Table~\ref{journal}. For our analysis we used the XMM-Newton SAS version 7.0, and the FTOOLS suite,  version 5.3.1.  For each observation, we filtered out intervals of high background (flaring), using the criteria recommended by the SAS team. We note that no flaring occurred during the  2001, January or 2005, November observations of NGC300 X-1. 

We then extracted pn and MOS 0.3--10 keV lightcurves and spectra from a circular region centred on the source, along with corresponding response files. Background regions were then chosen, and lightcurves and spectra were obtained from these regions for pn and MOS also. 

The pn and MOS source and background lightcurves were co-added, after careful synchronisation; XMM lightcurves are non-synchronised by default, leading to artificial variability if not treated properly \citep{bs07}. Background-subtracted lightcurves were analysed for variability, and PDS were made from the combined EPIC source lightcurves (background not subtracted). These lightcurves were averaged over several intervals of 1024 bins, with 5.2 s binning and geometric grouping; the sampled frequency range was $\sim$0.0002--0.1 Hz.

\begin{figure}[!t]
\resizebox{\hsize}{!}{\includegraphics[angle=270]{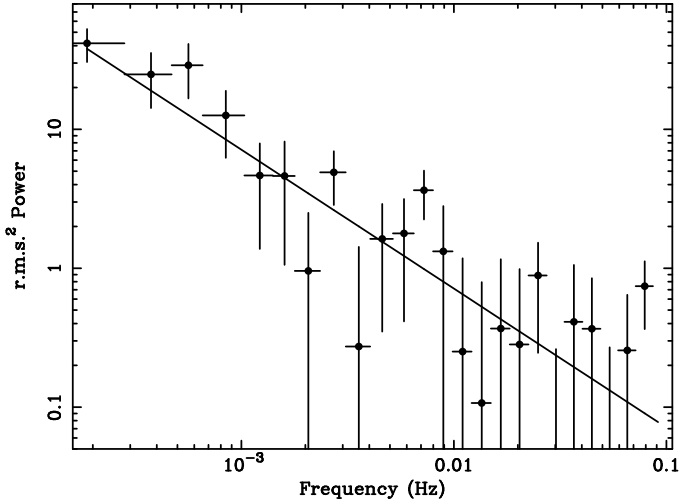}}
\caption{PDS of the combined pn+MOS 0.3--10 keV lightcurve of NGC300 X-1 from the 2005 November XMM-Newton observation (Obs. 4). The axes are log scaled, and the y axis is normalised to give r.m.s.$^{2}$ variability. The expected white noise level (7.48) is subtracted. A power law fit to the PDS is shown, with $\gamma$ = 1; the best fit $\chi^2$/d.o.f = 26/24, with a good fit probability of 0.35. Such a PDS is characteristic of a disc-accreting X-ray binary with a high accretion rate.
}\label{ngc300pds}
\end{figure}

\begin{figure}[!t]
\resizebox{\hsize}{!}{\includegraphics[angle=270]{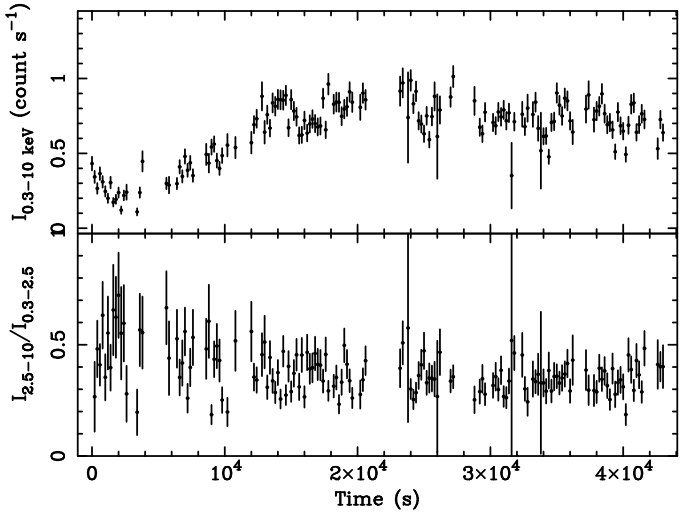}}
\caption{Combined EPIC lightcurve of IC10 X-1 from the 2003 XMM-Newton observation (top); the binning is 100 s. The lower panel shows the ratio of 2.5--10 keV photons to 0.3--2.5 keV photons (hardness ratio) for each time bin. The lightcurve is clearly variable, and there is evidence for variability in the hardness ratio.
}\label{ic10lc}
\end{figure}

\begin{figure}[!b]
\resizebox{\hsize}{!}{\includegraphics[angle=270]{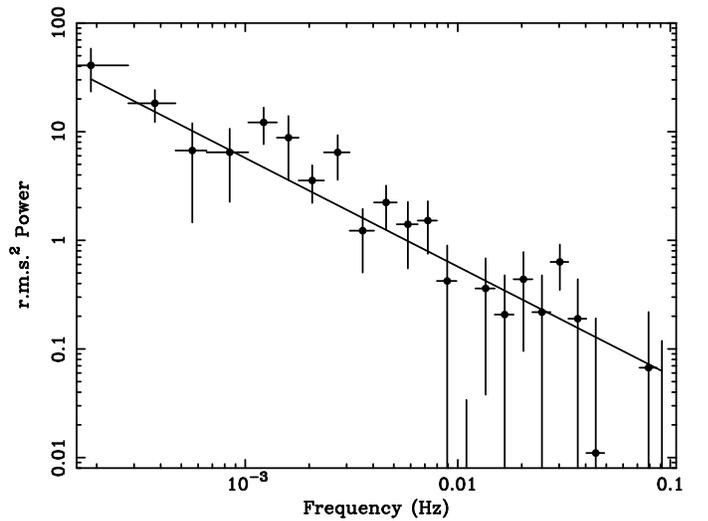}}
\caption{Combined EPIC PDS of IC10 X-1 from the 2003 XMM-Newton observation. The axes are log-scaled, and the y-axis is normalised to give r.m.s.$^2$ variability. The expected noise level (3.49) is subtracted. The best fit power law with $\gamma$ = 1  is shown, for $\chi^2$/dof = 35/24.
}\label{ic10pds}
\end{figure}

The source and background emission spectra from the two MOS cameras were combined  with the FTOOL mathpha to give MOS1+MOS2 source and background spectra, and the corresponding response matrices and ancillary response files were combined also, with addrmf and addarf. We modeled the pn and  combined MOS spectra simultaneously using XSPEC 11.3, with a constant of normalisation to account for differences in the pn and MOS responses.

\section{Results} 
\label{res}

 We discuss our analysis of the  variability and X-ray spectra  from NGC300 X-1 and IC10 X-1 in the following sections.

\subsection{Variability of NGC300 X-1 and IC10 X-1}

 The lightcurves from the four XMM-Newton observations of  NGC300 X-1 are well described by \citet{carp07a} and will not be discussed further here. Each observation { reveals strong variability}, with  PDS  that are well described by power laws with  $\gamma$ $\sim$1.
 We present  the 2005, November PDS in Fig.~\ref{ngc300pds}. The axes are log scaled, and normalised to give r.m.s.$^{2}$ variability; the expected noise is subtracted. We note that no background flaring occurred in this observation, so there is no question of the  PDS being an artefact of flaring or background filtering. This variability is certainly significant, as fitting  the PDS with zero power  yields a $\chi^2$ of 74 for 24 degrees of freedom (dof).
We present the r.m.s. variability and best fit $\gamma$ for each observation  of NGC300 X-1  in Table~\ref{300pds}. We note that the PDS from Obs. 3  is different to the PDS from the other observations; this difference is likely  due to the eclipse that occurs in Obs. 3 \citep[see ][]{carp07a}.

  We present the 0.3--10 keV, combined pn+MOS lightcurve of IC10 X-1 in the top panel of Fig.~\ref{ic10lc}; the background has been subtracted, and intervals of high background have been removed.  The lightcurve has 100 s resolution. The system is highly variable throughout the observation, with a large intensity dip near the beginning.  This dip may be intrinsic to the X-ray source, or due to an increase in line-of-sight absorption, such as an eclipse. The bottom panel shows the hardness ratio for each time bin, defined as the ratio of 2.5--10 keV counts to 0.3--2.5 keV counts. The hardness ratio is variable at the 3$\sigma$ level, and appears to be higher during the intensity dip at the start of the observation, hinting  at photo-electric absorption;  however, we cannot determine the phase of this intensity dip with the current orbital ephemeris. The variability is also consistent with stochastic variations in a source with a PDS described by $\gamma$ $\sim$1.

The PDS created from this lightcurve is presented in Fig.~\ref{ic10pds}; as before, the y-axis is normalised to show the r.m.s.$^{2}$ power, and the expected noise is subtracted. This PDS is also acceptably fitted  by a power law with $\gamma$ = 1, with $\chi^2$/dof = 35/24.  As with NGC300 X-1, the r.m.s. variability is greater than for Galactic LMXBs, at 18$\pm$2\%, excluding the large intensity dip.

   The PDS observed from NGC300 X-1 and IC10 X-1 are characteristic of  disc-fed XBs at high accretion rates, albeit with fractional r.m.s. variabilities $\sim$5--10 times higher than observed in Galactic XBs \citep[see e.g.][]{vdk95}.  However, they are also similar to the variability observed in certain wind-fed HMXBs. Hence, we must examine the emission spectra of NGC300 X-1 and IC10 X-1 if we are to determine their natures.

\begin{figure}[!t]
\resizebox{\hsize}{!}{\includegraphics[angle=270]{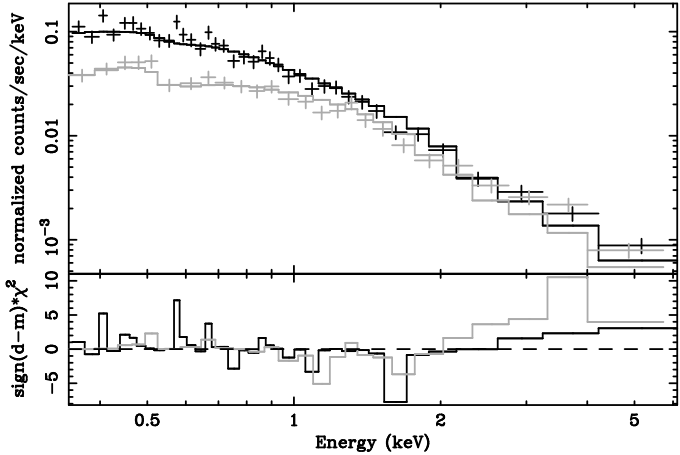}}
\caption{ {\em Top panel:} Best fit model to the Obs. 1 pn (black) and combined MOS (grey) 0.3--10 keV spectra of NGC300 X-1 using a power law with $\Gamma$ = 2.44 and a 0.95 keV Gaussian line with width 0.07 keV;  $\chi^2$/dof = 100/60, which is unacceptable {\em Bottom panel:} $\chi^2$ residuals, showing a hard excess; this shows that  a flatter power law and/or a second emission component is required. The spectrum of NGC300 X-1 from Obs. 1 is certainly different from the model obtained by \citet{carp07a} when fitting spectra from all four observations simultaneously.}\label{ob1300spec}
\end{figure}

\subsection{ Modeling the spectra of NGC300 X-1 and IC10 X-1}

\subsubsection{NGC300 X-1}
 \citet{carp07a} modeled the spectra from the four observations of NGC300 X-1 together. They used   an absorbed, two-component model, consisting of a power law  { with photon index $\Gamma$, i.e. $N(E)\propto E^{-\Gamma}$, where $N$ is the number of photons at energy $E$}, and an emission line. { They imposed identical values for $\Gamma$, line energy and line width for the four observations, with these values free to vary}; the absorption and normalizations for each component were individually varied for each observation. For their best model, $\Gamma$ = 2.44$\pm$0.03 ,  the line energy was  0.94$\pm$0.01 keV, with a width of 0.07$\pm$0.01 keV. They quote a reduced  $\chi^2$ of  1.15 for 1033 dof; however, this equates to $\chi^2$/dof = 1188/1033, which is rejected at a 99.95\% level and is therefore unacceptable. We therefore examined the spectra of NGC300 X-1 more carefully.

  For each observation of NGC300 X-1, we modeled the pn and combined MOS spectra  simultaneously, with each model consisting of a power law + emission line, suffering absorption and  including a constant of normalization to account for differences in pn and MOS calibration. The resulting best fits are shown in Table~\ref{specfits}. We found the spectra from Obs. 2--4 to be consistent with the model reported by \citet{carp07a}. However, the best fit to the Obs. 1 spectra yielded $\chi^2$/dof = 100/60, an unacceptable fit. We show this best fit in Fig.~\ref{ob1300spec}, along with $\chi^2$ residuals. We see that the model systematically underestimates the emission at higher energies, suggestive of an additional emission component. Hence, we fitted all NGC300 X-1 spectra with absorbed blackbody  + power law models and also absorbed  disc blackbody + power law models, commonly observed in NS and BH XBs respectively. The best fits for these models are also provided in Table~\ref{specfits}; 0.3--10 keV luminosities are provided, assuming a distance of 1.88 Mpc \citep{gps05}.
  We see that both two-component models provided good fits to the Obs. 1 and Obs. 3 spectra. The Obs. 3 spectrum is well described by all three models; however, the resulting luminosities differ by factor of two ($\sim$5$\sigma$); hence the { nature of the Obs. 3. emission spectrum is uncertain}. Neither the Obs. 2 nor Obs. 4 spectra were successfully described by any thermal + power law  model; hence  NGC3000 X-1 exhibits at least two spectral states: one exhibited in Obs. 1 and the other exhibited in Obs. 2 and 4.

\begin{table}[!b]
 \centering
  \caption{ Fits to the 0.3--10 keV PDS of NGC300 X-1 from Obs. 1--4. We give the observation number, as well the best fit spectral index with the  $\chi^2$/dof. We also give the $\chi^2$/dof for zero power. Finally we give the fractional r.m.s. variability. We note that the PDS for Obs. 3 is more complex than for the other observations, due to an eclipse.  }\label{300pds}
  \begin{tabular}{ccccccc}
  \noalign{\smallskip}
  \hline
  \noalign{\smallskip}
  Obs. &  $\gamma$  & $\chi^2$/dof & $\chi^2$/dof (0) &   Frac. r.m.s. (\%) \\
\noalign{\smallskip}
 \hline
\noalign{\smallskip}  
 Obs. 1& 1.30(12) & 17/23 & 57/25 & 50(3) \\
 Obs. 2& 1.3(2) & 27/23 & 66/25 & 44(2) \\
 Obs. 3&  1.8(2) & 56/23 & 91/25 & 38(2) \\
 Obs. 4 & 1.05(11) & 23/23 & 74/23 & 22(2) \\
\noalign{\smallskip}
\hline
\noalign{\smallskip}
\end{tabular}
\end{table}

\begin{figure}[!t]
\resizebox{\hsize}{!}{\includegraphics[angle=270]{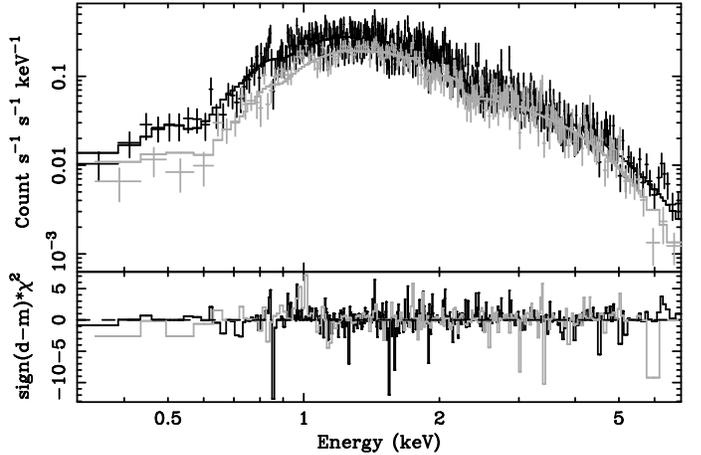}}
\caption{ Simultaneous fits to the pn (black) and combined MOS (grey) spectra of IC10 X-1 from the 2003 XMM-Newton observation with the best fit emission model; the first 20 ks of the observation was excluded.  The top panel shows the spectra modeled with a power law with photon index 3.5$\pm$0.2 and a 0.86$\pm$0.03 keV keV blackbody, suffering absorption equivalent to 8.6$\pm$0.4$\times 10^{21}$ H atom cm$^{-2}$, with $\chi^2$= 849 for 840 degrees of freedom (good fit probability 0.4).  The bottom panel shows the $\chi^2$ residuals. 
}\label{ic10spec}
\end{figure}

\begin{table*}[!t]
 \centering
  \caption{ Fits to the spectra of NGC300 X-1 from Obs. 1--4,  and of IC10 X-1. The first model is an absorbed power law plus Gaussian emission line, as favoured by Carpano et al. (2007a). For this model, we give absorption, normalised to 10$^{22}$ H atom cm$^{-2}$, spectral index, line energy and width (in keV), and $\chi^2$/dof,  with good fit probability in square brackets. We only apply the Gaussian to NGC300 X-1. We also give the 0.3--10 keV luminosity normalised to 10$^{37}$ erg  s$^{-1}$.  We next provide the best fits for a disk blackbody + power law model that is often used when describing the high state of disc-accreting BH XBs. We give the absorption, temperature of the disk blackbody, spectral index of the power law component and $\chi^2$. We then give 0.3--10 keV luminosity as before, and the fractional contribution of the power law model. }\label{specfits}
  \begin{tabular}{ccccccc}
  \noalign{\smallskip}
  \hline
  \noalign{\smallskip}
        &           &           &    NGC300 X-1 &  & & IC10 X-1  \\
  Model &  Parameter  & Obs. 1  &  Obs. 2 & Obs. 3 & Obs. 4 &  \\
\noalign{\smallskip}
 \hline
\noalign{\smallskip}  
PL + Line & $n_{\rm H}$ & 0.036  & 0.067(12)& 0.052(9) & 0.07(2) & 0.75(3)\\
 & $\Gamma$ &  2.46(6) & 2.44(7) & 2.44 & 2.53(8) & 2.56(6) \\
 & $E_{\rm Line}$ & 0.95 &  0.93(2) & 0.95(2) & 0.95(2) & $\dots$\\
 & $\sigma$ & 0.07 &  0.08(2)  & 0.07 & 0.06(3)  & $\dots$ \\
 & $\chi^2$/dof & 100/60 [9E-4] & 149/154 [0.60] & 50/60 [0.82]& 117/145 [0.05] & 794/652 [1E-4]\\
 & $L_{\rm 0.3-10 keV}$ & 11.5(13) & 28.4(12) & 20(2) & 34.8(16) & 39(3)\vspace{0.1in}\\

BB+PL & $n_{\rm H}$ & 0.11(4) & 0.130(12) & 0.22(5) & 0.17(3)& 0.92(10)\\
 & k$T$ & 1.02(14) & 1.02(17) & 0.92(12) & 1.08(15) & 0.82(3) \\
 & $\Gamma$ & 3.41(6) & 3.01(16) & 3.8(3) & 3.26(6) & 3.9(5)\\
 & $\chi^2$/dof & 52/58 [0.67] & 211/155 [2E-3] & 59/58[0.43] & 221/148[1.0E-4] & 697/650 [0.09]\\
&$L_{\rm 0.3-10 keV}$ & 16.9(17) & 39(2) & 49(6) & 59(3) & 91(19)\\
&Frac. PL & 0.80(11) & 0.93(14) & 0.91(18) & 0.92(6) & 0.93(3)\vspace{0.1in}\\

DISKBB+PL & $n_{\rm H}$ & 0.12(4) & 0.13(4) & 0.22(9) & 0.17(4) & 0.84(4)\\
 & k$T_{\rm in}$ & 1.8(4) & 1.9(6) & 1.5(4) & 2.0(4) & 1.19(6)\\
 & $\Gamma$ & 3.6(7) & 3.1(4) & 4.1(7) & 3.4(3) & 3.9 \\
 & $\chi^2$/dof & 56/58 [0.54] & 213/155 [1.3E-3] & 62/58[0.33] & 226/148[3.0E-5] & 710/651 [0.05]\\
&$L_{\rm 0.3-10 keV}$ & 19(3) & 33(25) & 49(6) & 53(5) & 72(9)\\
&Frac. PL & 0.89(17) & 0.6(7) & 0.91(18) & 0.93(11) & 0.86(19)\\
\noalign{\smallskip}
\hline
\noalign{\smallskip}
\end{tabular}
\end{table*}

Finally, we  fitted the spectra with { the  hard power law emission that is characteristic of  Bondi-Hoyle accretion onto a neutron star}. Our literature review shows that the wind-accreting HMXBs with PDS similar to those observed from NGC300 X-1 all have spectra that are well described by a power law with  $\Gamma$ $\le$1.8. Hence, we modeled the spectra from each observation with power law models, restricting $\Gamma$ to be $<$1.8.  We found the best fit $\chi^2$/dof to be $\ga$5 in each case. Hence NGC300 X-1 is unlikely to be powered by Bondi-Hoyle accretion onto a neutron star.

\subsubsection{IC10 X-1}

We first attempted to simultaneously fit the pn and combined MOS spectra of IC10 X-1  with various emission models, but failed to obtain a good fit. This is likely to be due to the fact that IC10 X-1 varies in colour and intensity during the observation, mixing the spectral states so that a physical fit is impossible.

 In light of this, we excluded the first 20 ks of the observation, where the intensity and colour of IC10 X-1 is unstable. We modeled the spectrum of the remaining $\sim$22 ks with  power law, blackbody + power law, and disc blackbody + power law emission models, all suffering absorption. The best fits are presented in Table~\ref{specfits}; quoted luminosities assume a distance of 740 kpc \citep{dem04}. The disc blackbody + power law model favoured an unreasonably high $\Gamma$, although $\Gamma$ was not well constrained. Hence, we fixed it to 3.9, the best fit $\Gamma$ for the blackbody + power law model.  No simple power law model was able to fit, but both two-component models provided acceptable fits. However, since we know that IC10 X-1 contains a BH, we favour the disc blackbody + power law model.  We present the best fit to the IC10 X-1 spectrum with the disc blackbody + power law model in Fig.~\ref{ic10spec}.

\citet{wang05} modeled pn and MOS spectra from the entire XMM-Newton observation, mixing spectral states  and thereby producing a spectrum that is not representative of either state. As a result, they required a metal abundance $<$0.01 Solar to achieve good fits; they remark that such a low metallicity is probably not physical, but attribute it to poor calibration at low energies rather than to spectral mixing. Their preferred model is a self-consistent Comptonised multi-colour disc blackbody where an accretion disc emits a multi-temperature disc blackbody, which is  Comptonised by a spherically symmetric corona of hot electrons around the disc. Hence \citet{wang05} also favour a non-thermal emission model for IC10 X-1. { They estimated the mass of the black hole by assuming that the inner disc radius corresponds to the innermost stable orbit, after correcting for relativistic effects. They obtained a BH mass range of  $\sim$4 M$_{\odot}$ for a stationary BH, up to $\sim$30 M$_{\odot}$ for a BH with extreme spin; this range  overlaps the mass range found by \citet{sf08}. }

\section{Discussion}
\label{discuss}

 We have examined the variability and emission spectra of one BH+WR candidate (NGC300 X-1), and the newly  confirmed BH+WR system IC10 X-1. The 0.3--10 keV  spectrum of Obs. 1 of NGC300 X-1 is strikingly similar to that of IC10 X-1. This lends strong support to the case for NGC300 X-1 being a BH+WR system. Both systems exhibit variability that scales  with frequency, $\nu$, as $\nu^{-1}$, identified with BH XBs in the high state. However, their 0.3--10 keV emission spectra appear to be $\sim$90\% non-thermal, while the canonical high state BH spectrum is thermally dominated. We therefore suggest that these systems exist in a previously unrecognised state.

\subsection{Accretion scenarios}

We next discuss different accretion scenarios. IC10 X-1 has been confirmed as a WR+BH binary, hence we first discuss wind and disc accretion onto a  BH. NGC300 X-1 has an unknown accretor, and may contain a neutron star; therefore, we also discuss wind and disc accretion onto a NS.

Since the emission from Bondi-Hoyle accreting neutron stars appears to originate on or near the surface of the neutron star, it is unclear what X-ray emission could be expected from Bondi-Hoyle accretion onto a black hole. Indeed, there are no known black hole Be XBs, and all known black hole supergiant XBs exhibit  accretion discs.
Furthermore, the modeling of Bondi-Hoyle accretion  onto black holes   requires relativistic, three-dimensional magnetohydrodynamical modeling \citep[see. e.g.][ and references within]{font99}. These models have not yet produced unambiguous predictions for the variability or emission mechanisms expected from such systems.  Hence, there are no observational or theoretical constraints on the spectral shape or variability exhibited by HMXBs powered by Bondi-Hoyle accretion onto a BH.

  The short orbital periods  ($\sim$30 hr) and high X-ray luminosities of NGC300 X-1 and IC10 X-1 resemble those of disc-accreting SG HMXBs \citep{kap04}; we therefore consider them likely  disc-accretors also.
 \citet{prest07} calculated the range of orbital periods that permit Roche lobe overflow for IC10 X-1. They obtained periods of  $\sim$2.5--3 hr, and rejected disc accretion for IC10 X-1  (this also applies to  NGC300 X-1, as it  has a similar orbital period to IC10 X-1).
  However, it is entirely possible for the winds of the Wolf-Rayet stars to power disc accretion.  The three known black hole HMXBs Cygnus X-1, LMC X-1 and LMC X-3  are all disc-accreting, with orbital periods of 5.60 d,  4.22 d and 1.70 d respectively \citep[see e.g.][and references within]{lvv95}.  Of these, only LMC X-3 is thought to be  Roche lobe filling.   NGC300 X-1 and IC10 X-1 have shorter orbital periods than any of these systems, so disc accretion onto a BH in these systems is entirely plausible.

 As discussed in Sect.~3, our spectral modeling of NGC300 X-1 allows us to rule out Bondi-Hoyle (wind) accretion onto a NS, as power law spectral fits with $\Gamma$ $\le$ 1.8 yield $\chi^2$/dof $>$5 for all observations. However, the emission and variability of NGC300 X-1 are in keeping with a disc-fed NS XB. The donor star may be the WR star, in which case, the disc would likely be wind-fed. Alternatively, NGC300 X-1 could be a bright NS LMXB which is merely coincident with the WR.

\subsection{A new black hole state?}

 We prefer disc accretion scenarios for IC10 X-1 and NGC300 X-1, although there is no independent evidence for a disc, such as radio jets. All known BH XBs are disc accreting, yet the properties exhibited by IC10 X-1 and NGC300 X-1 are as yet unrecognised. Their observed emission spectra most closely resemble the very high (steep power law, SPL) state \citep{vdk95, mr03}. However,  the PDS of sources in the SPL state are characterised by broken power laws, where $\gamma$ changes from $\sim$0 at low frequencies to $\sim$1 at high frequencies; often  quasi-periodic oscillations (QPOs) are observed \citep{mr03}.

 We note that Cyg X-3,  the other WR+co candidate,  exhibits similar timing behaviour to NGC300 X-1 and IC10 X-1. Some believe that scattering in the wind suppresses high frequency  variability \citep[see e.g.][]{kit92}, possibly masking a { SPL PDS}. Hence, one might wonder if  NGC300 X-1 (in Observation 1) or IC10 X-1 are disguising SPL states. However, the r.m.s. variability of NGC300 X-1 was already 44\% in Observation 1, while the r.m.s. variability of IC10 X-1 was 18\%. Hence they are already considerably more variable than BH LMXBs in the very high state \citep{vdk95}, making this hypothesis unlikely.

 Instead, we believe that the observed behaviour maybe due to the status of the corona.
In NS LMXBs, the self-irradiation of the disc is more efficient than in BH LMXBs, hence NS LMXBs have hot, stable discs, while  all BH LMXBs are transient \citep{kk97}; they spend the majority of their time in a quiescent state, with outbursts generally lasting a few months \citep[see e.g.][]{mr03}.   During the rise of the outburst, a BH LMXB will go from a low accretion rate, hard state with associated radio jets, to a high accretion rate, soft state with no jets \citep[see e.g.][ and references within]{hb05}.
It is therefore possible that the source of the hard, non-thermal emission, and also of the jets in BH LMXBs (e.g. the corona) is ejected during this transition; indeed \citet{gal04} witness a large radio flare that coincided with a hard to soft transition in GX 339$-$4.

By contrast, the three known Galactic BH HMXBs are all persistently bright. The HMXB accretion discs are expected to be small, and hence will be more easily kept hot at the outer edge, making them stable. 

 Therefore, the behaviour of IC10 X-1 and NGC300 X-1 may be explained by stable disc accretion.
 A small fraction of the  $\sim 10^{-5}-10^{-4}$ M$_{\odot}$ yr$^{-1}$ wind from a WR companion leaving the vicinity of the  L1 point could provide enough mass, with sufficient angular momentum, to sustain an accretion disc that is persistently in the high state. In this case, the corona could be retained, and the emission would then be dominated by a non-thermal component due to inverse-Comptonisation of cool photons on hot electrons, as seen in NS XBs at high accretion rate.

 Observations of the Galactic BH HMXBs support this hypothesis.
 LMC X-1 has never been observed in the low/hard state \citep[see e.g.][]{wilms01,yao05}.  \citet{now01} report ``high state'' (i.e. power $\propto$ $\nu^{-1}$) PDS from RXTE observations of LMC X-1, while \citet{wilms01} find its emission spectrum to be dominated by a  power law with photon index $\ga$3, with a 0.8--1.1 keV blackbody component.  This emission spectrum is remarkably similar to our best fit models for NGC300 X-1 in Obs. 1 , and to  IC10 X-1.  In contrast, Cyg X-1 and LMC X-3 exhibit transitions between the low/hard and high states, and exhibit canonical spectra \citep[and references within]{mr03}.

\subsection{Comparison with ULXs}

 It is interesting to compare these systems with the ULXs, as many ULXs are thought to be HMXBs. { In their review of black hole masses and spectral states in ULXs, \citet{sk08} find that only $\sim$10\% of sources with luminosities above 10$^{39}$ erg s$^{-1}$ exhibit high/soft (i.e. thermally dominated) spectra, with  almost all ULXs  instead dominated by a broad ``power law-like'' component that contributes $\sim$90\% of the flux. Such characteristics are remarkably similar to those of NGC300 X-1 and IC10 X-1, where the power law contributes $\sim$80--90\% of the flux. It is therefore possible that these ULXs could also be peresistently bright with stable coronae. It is therefore instructive to study the   spectral and, crucially, timing properties of ULXs.

}

 \citet{wmr06} conducted a survey of all point sources $>$10$^{38}$ erg s$^{-1}$ in XMM-Newton observations of 32 nearby galaxies. They modeled the spectra of $\sim$100 X-ray sources, each with $\ga$400 counts. Ten sources exhibited spectra consistent with NGC300 X-1 in Obs. 2 and 4, including three  ULXs. A further 23 sources exhibited spectra consistent with Obs. 1 of NGC300 X-1  and IC10 X-1, including 10 ULXs. { Hence the processes in NGC300 X-1 and IC10 X-1 could also occur in some ULXs.

While most of the known ULXs are too faint to allow fruitful analysis of the PDS, several of the nearest ULXs are as bright as, or brighter than, NGC300 X-1 and IC10 X-1. A full analysis of the timing properties of nearby ULXs is beyond the scope of this work, and is the subject of a forthcoming paper.  However, we conducted a brief literature review of ULX variability. Several ULXs exhibit QPOs, e.g.  NGC5408 X-1 \citep{smw07}, M82 X-1 \citep{sm03} and HoIX  X-1 \citep{dgr06}. \citet{smw07} the NGC5408 X-1 PDS to be characterised by a broken power law; this PDS, along with the emission spectrum (dominated by a power law with $\Gamma$ $\sim$2.6) is reminiscent of the SPL state of Galactic black holes, unlike NGC300 X-1 and IC10 X-1. The published PDS of M82 X-1 and HoIX X-1 do not extend to sufficiently low frequencies to determine whether they are also in the SPL state. We note that the broken power law PDS and power law PDS identified by  \citet{crop04} in NGC4559 X-7 and NGC4559 X-10 are both artificial, caused by improper treatment of non-sychronised lightcurves \citep{bs07}. 

}

\section{Conclusions}
\label{conc}

 IC10 X-1 is the first confirmed BH+WR system, and NGC300 X-1 is thought to be one also. They are known to have strikingly similar X-ray periods and luminosities, so we studied  them in detail.
We have examined for the first time the PDS of NGC300 X-1, in four XMM-Newton observations, and also of IC10 X-1, in one XMM-Newton observation. We find that they are  well described by a simple power law with spectral index $\simeq$1, over the 0.002--0.1 Hz range; such variability is characteristic of disc-accreting XBs in their high state,  or of some HMXB pulsars that are powered by Bondi-Hoyle accretion.  We also find  a striking resemblance between the Obs. 1 NGC300 X-1 spectrum and the IC10 X-1 spectrum. Since IC10 X-1 is a confirmed WR+BH system, our results provide strong support for NGC300 X-1 being one too.

 We have considered four accretion scenarios:  disc accretion or  Bondi-Hoyle accretion,  onto a BH or NS. We favour disc accretion onto a BH for both systems, but cannot rule out Bond-Hoyle accretion onto a BH. The observed X-ray spectra of NGC300 X-1 allowed  us to rule out Bondi-Hoyle accretion onto a NS,  but do not exclude disc accretion onto a NS+WR binary, or onto a NS LMXB that is merely coincident with the WR.

If  both NGC300 X-1 and IC10 X-1  are indeed WR+BH binaries, then they are spectrally distinct from the known BH XBs. We propose that this difference may be due to these systems being persistently in the high state, allowing them to keep their disc corona,  like LMC X-1; contrariwise, BH LMXBs enter the high state only in violent outbursts, where the corona may be ejected. 
 NGC300 X-1 and IC10 X-1 (and possibly LMC X-1) would then comprise a new class of BH binary, defined by their 
high mass transfer rates and consequent stable accretion disc configurations. { This stable disc scenario may also explain why only $\sim$10\% of ULXs exist in the thermally-dominated high soft state. A fair comparison between the ULXs and NGC300 X-1 and IC10 X-1 requires detailed analyis of the variability of ULXs in addition to spectral analysis; we will carry this out in a future paper.}

\section*{Acknowledgments}

PDS were fitted using  fitpowspec, provided by P.J. Humphrey. Astrophysics at the Open University is funded by a  STFC (formerly PPARC) rolling grant. We thank Paul Crowther and Andrea Prestwich for communicating results prior to publication.  We are also grateful to the anonymous referee for a very constructive review of the paper.

\bibliographystyle{aa}
\bibliography{m31}

\begin{thebibliography}{39}
\expandafter\ifx\csname natexlab\endcsname\relax\def\natexlab#1{#1}\fi

\bibitem[{{Barnard} {et~al.}(2007){Barnard}, {Trudolyubov}, {Kolb}, {Haswell},
  {Osborne}, \& {Priedhorsky}}]{bs07}
{Barnard}, R., {Trudolyubov}, S., {Kolb}, U.~C., {et~al.} 2007, \aap, 469, 875

\bibitem[{{Bauer} \& {Brandt}(2004)}]{bb04}
{Bauer}, F.~E. \& {Brandt}, W.~N. 2004, \apjl, 601, L67

\bibitem[{{Belloni} \& {Hasinger}(1990)}]{bh90}
{Belloni}, T. \& {Hasinger}, G. 1990, A\&A, 227, L33

\bibitem[{{Bondi} \& {Hoyle}(1944)}]{bh44}
{Bondi}, H. \& {Hoyle}, F. 1944, \mnras, 104, 273

\bibitem[{{Carpano} {et~al.}(2007){Carpano}, {Pollock}, {Wilms}, {Ehle}, \&
  {Schirmer}}]{carp07a}
{Carpano}, S., {Pollock}, A.~M.~T., {Wilms}, J., {Ehle}, M., \& {Schirmer}, M.
  2007, \aap, 461, L9

\bibitem[{{Clark} \& {Crowther}(2004)}]{cc04}
{Clark}, J.~S. \& {Crowther}, P.~A. 2004, \aap, 414, L45

\bibitem[{{Cropper} {et~al.}(2004){Cropper}, {Soria}, {Mushotzky}, {Wu},
  {Markwardt}, \& {Pakull}}]{crop04}
{Cropper}, M., {Soria}, R., {Mushotzky}, R.~F., {et~al.} 2004, \mnras, 349, 39

\bibitem[{{Crowther}(2007)}]{review}
{Crowther}, P.~A. 2007, \araa, 45, 177

\bibitem[{{Crowther} {et~al.}(2007){Crowther}, {Carpano}, {Hadfield}, \&
  {Pollock}}]{cch07}
{Crowther}, P.~A., {Carpano}, S., {Hadfield}, L.~J., \& {Pollock}, A.~M.~T.
  2007, \aap, 469, L31

\bibitem[{{Demers} {et~al.}(2004){Demers}, {Battinelli}, \& {Letarte}}]{dem04}
{Demers}, S., {Battinelli}, P., \& {Letarte}, B. 2004, \aap, 424, 125

\bibitem[{{Dewangan} {et~al.}(2006){Dewangan}, {Griffiths}, \& {Rao}}]{dgr06}
{Dewangan}, G.~C., {Griffiths}, R.~E., \& {Rao}, A.~R. 2006, \apjl, 641, L125

\bibitem[{{Done} \& {Gierli{\'n}ski}(2004)}]{don04}
{Done}, C. \& {Gierli{\'n}ski}, M. 2004, Progress of Theoretical Physics
  Supplement, 155, 9

\bibitem[{{Font} {et~al.}(1999){Font}, {Ib{\'a}{\~n}ez}, \&
  {Papadopoulos}}]{font99}
{Font}, J.~A., {Ib{\'a}{\~n}ez}, J.~M., \& {Papadopoulos}, P. 1999, \mnras,
  305, 920

\bibitem[{{Gallo} {et~al.}(2004){Gallo}, {Corbel}, {Fender}, {Maccarone}, \&
  {Tzioumis}}]{gal04}
{Gallo}, E., {Corbel}, S., {Fender}, R.~P., {Maccarone}, T.~J., \& {Tzioumis},
  A.~K. 2004, \mnras, 347, L52

\bibitem[{{Gieren} {et~al.}(2005){Gieren}, {Pietrzy{\'n}ski}, {Soszy{\'n}ski},
  {Bresolin}, {Kudritzki}, {Minniti}, \& {Storm}}]{gps05}
{Gieren}, W., {Pietrzy{\'n}ski}, G., {Soszy{\'n}ski}, I., {et~al.} 2005, \apj,
  628, 695

\bibitem[{{Homan} \& {Belloni}(2005)}]{hb05}
{Homan}, J. \& {Belloni}, T. 2005, \apss, 300, 107

\bibitem[{{Kaper} {et~al.}(2004){Kaper}, {van der Meer}, \& {Tijani}}]{kap04}
{Kaper}, L., {van der Meer}, A., \& {Tijani}, A.~H. 2004, in Revista Mexicana
  de Astronomia y Astrofisica Conference Series, Vol.~21, Revista Mexicana de
  Astronomia y Astrofisica Conference Series, ed. C.~{Allen} \& C.~{Scarfe},
  128--131

\bibitem[{{King} {et~al.}(1997){King}, {Kolb}, \& {Szuszkiewicz}}]{kk97}
{King}, A.~R., {Kolb}, U., \& {Szuszkiewicz}, E. 1997, \apj, 488, 89

\bibitem[{{Kitamoto} {et~al.}(1992){Kitamoto}, {Mizobuchi}, {Yamashita}, \&
  {Nakamura}}]{kit92}
{Kitamoto}, S., {Mizobuchi}, S., {Yamashita}, K., \& {Nakamura}, H. 1992, \apj,
  384, 263

\bibitem[{{Lewin} {et~al.}(1995){Lewin}, {van Paradijs}, \& {van den
  Heuvel}}]{lvv95}
{Lewin}, W.~H.~G., {van Paradijs}, J., \& {van den Heuvel}, E.~p.~J. 1995,
  {X-ray Binaries} ({Cambridge University Press})

\bibitem[{{Liu} {et~al.}(2005){Liu}, {van Paradijs}, \& {van den
  Heuvel}}]{lvv05}
{Liu}, Q.~Z., {van Paradijs}, J., \& {van den Heuvel}, E.~P.~J. 2005, \aap,
  442, 1135

\bibitem[{{Liu} {et~al.}(2007){Liu}, {van Paradijs}, \& {van den
  Heuvel}}]{lvv07}
---. 2007, \aap, 469, 807

\bibitem[{{Lommen} {et~al.}(2005){Lommen}, {Yungelson}, {van den Heuvel},
  {Nelemans}, \& {Portegies Zwart}}]{lom05}
{Lommen}, D., {Yungelson}, L., {van den Heuvel}, E., {Nelemans}, G., \&
  {Portegies Zwart}, S. 2005, \aap, 443, 231

\bibitem[{{McClintock} \& {Remillard}(2006)}]{mr03}
{McClintock}, J.~E. \& {Remillard}, R.~A. 2006, {Black hole binaries} (Compact
  stellar X-ray sources), 157--213

\bibitem[{{Nagase}(1989)}]{nag89}
{Nagase}, F. 1989, \pasj, 41, 1

\bibitem[{{Nowak} {et~al.}(2001){Nowak}, {Wilms}, {Heindl}, {Pottschmidt},
  {Dove}, \& {Begelman}}]{now01}
{Nowak}, M.~A., {Wilms}, J., {Heindl}, W.~A., {et~al.} 2001, \mnras, 320, 316

\bibitem[{{Okazaki} \& {Negueruela}(2001)}]{on01}
{Okazaki}, A.~T. \& {Negueruela}, I. 2001, \aap, 377, 161

\bibitem[{{Prestwich} {et~al.}(2007){Prestwich}, {Kilgard}, {Crowther},
  {Carpano}, {Pollock}, {Zezas}, {Saar}, {Roberts}, \& {Ward}}]{prest07}
{Prestwich}, A.~H., {Kilgard}, R., {Crowther}, P.~A., {et~al.} 2007, ApJL, 669,
  L21

\bibitem[{{Silverman} \& {Filippenko}(2008)}]{sf08}
{Silverman}, J.~M. \& {Filippenko}, A.~V. 2008, ApJL, 678, L17

\bibitem[{{Soria} \& {Kuncic}(2008)}]{sk08}
{Soria}, R. \& {Kuncic}, Z. 2008, Advances in Space Research, 42, 517

\bibitem[{{Strohmayer} \& {Mushotzky}(2003)}]{sm03}
{Strohmayer}, T.~E. \& {Mushotzky}, R.~F. 2003, \apjl, 586, L61

\bibitem[{{Strohmayer} {et~al.}(2007){Strohmayer}, {Mushotzky}, {Winter},
  {Soria}, {Uttley}, \& {Cropper}}]{smw07}
{Strohmayer}, T.~E., {Mushotzky}, R.~F., {Winter}, L., {et~al.} 2007, \apj,
  660, 580

\bibitem[{{van der Klis}(1994)}]{vdk94}
{van der Klis}, M. 1994, ApJs, 92, 511

\bibitem[{{van der Klis}(1995)}]{vdk95}
---. 1995, {X-ray Binaries} ({Cambridge University Press}), 256--307

\bibitem[{{Wang} {et~al.}(2005){Wang}, {Whitaker}, \& {Williams}}]{wang05}
{Wang}, Q.~D., {Whitaker}, K.~E., \& {Williams}, R. 2005, \mnras, 362, 1065

\bibitem[{{White} {et~al.}(1995){White}, {Nagase}, \& {Parmar}}]{wnp95}
{White}, N.~E., {Nagase}, F., \& {Parmar}, A.~N. 1995, {X-ray Binaries}
  ({Cambridge University Press})

\bibitem[{{Wilms} {et~al.}(2001){Wilms}, {Nowak}, {Pottschmidt}, {Heindl},
  {Dove}, \& {Begelman}}]{wilms01}
{Wilms}, J., {Nowak}, M.~A., {Pottschmidt}, K., {et~al.} 2001, MNRAS, 320, 327

\bibitem[{{Winter} {et~al.}(2006){Winter}, {Mushotzky}, \& {Reynolds}}]{wmr06}
{Winter}, L.~M., {Mushotzky}, R.~F., \& {Reynolds}, C.~S. 2006, \apj, 649, 730

\bibitem[{{Yao} {et~al.}(2005){Yao}, {Wang}, \& {Nan Zhang}}]{yao05}
{Yao}, Y., {Wang}, Q.~D., \& {Nan Zhang}, S. 2005, \mnras, 362, 229

\end{thebibliography}

\end{document}